%% file: sparql-autocompletion.tex
\documentclass[sigconf, nonacm]{acmart}

\usepackage{placeins}[subsection] 
\usepackage{booktabs} 
\usepackage{array,tabularx}
\usepackage{enumitem}
\usepackage{csquotes}
\usepackage{graphicx} \pdfcompresslevel=9
\usepackage{hyperref}
\usepackage{colortbl}
\usepackage{tikz}
\usepackage{bm}

\newcolumntype{C}[1]{>{\centering\arraybackslash}m{#1}}
\newcolumntype{Y}{>{\centering\arraybackslash}X}

\usepackage{xcolor}
\definecolor{darkblue}{rgb}{0.0, 0.0, 0.7}

\newenvironment{sparqlkb}
{\it\vspace{.3mm}\begin{tabbing}\hspace*{32mm}\=\hspace*{22mm}\=\kill}
{\end{tabbing}\vspace{.3mm}\noindent\ignorespacesafterend}
\newenvironment{sparqlkb2}
{\it\vspace{.3mm}\begin{tabbing}\hspace*{25mm}\=\hspace*{25mm}\=\kill}
{\end{tabbing}\vspace{.3mm}\noindent\ignorespacesafterend}
\newenvironment{sparqlq}
{\it\vspace{.3mm}\begin{tabbing}\hspace*{3mm}\=\hspace*{14mm}\=\hspace*{24mm}
\=\hspace*{18mm}\=\kill}
{\end{tabbing}\vspace{0mm}\noindent\ignorespacesafterend}
\newenvironment{sparqlqq}
{\it\vspace{.3mm}\renewcommand{\baselinestretch}{1.1}\normalsize\begin{tabbing}\hspace*{7mm}\=\hspace*{3mm}\=\hspace*{5mm}
\=\hspace*{3mm}\=\kill}
{\end{tabbing}\vspace{.3mm}\noindent\ignorespacesafterend\renewcommand{\baselinestretch}{1.0}\normalsize}

\def\sparqlemph#1{#1}
\def\SELECT{\sparqlemph{SELECT}}
\def\DISTINCT{\sparqlemph{DISTINCT}}
\def\FROM{\sparqlemph{FROM}}
\def\WHERE{\sparqlemph{WHERE}}
\def\LIMIT{\sparqlemph{LIMIT}}
\def\ORDERBY{\sparqlemph{ORDER BY}}
\def\GROUPBY{\sparqlemph{GROUP BY}}
\def\DESC{\sparqlemph{DESC}}

\def\FILTER{\sparqlemph{FILTER}}
\def\LANG{\sparqlemph{LANG}}
\def\REGEX{\sparqlemph{REGEX}}
\def\COUNT{\sparqlemph{COUNT}}
\def\AS{\sparqlemph{AS}}
\def\OPTIONAL{\sparqlemph{OPTIONAL}}
\def\UNION{\sparqlemph{UNION}}
\def\MINUS{\sparqlemph{MINUS}}
\def\BIND{\sparqlemph{BIND}}
\def\STR{\sparqlemph{STR}}
\def\SAMPLE{\sparqlemph{SAMPLE}}
\def\cursor{\hspace{0.3mm}\underline{\hspace{1.8mm}}}

\def\KS#1{\mathrm{KS}_{#1}}

\def\MRR#1{\mathrm{MRR}_{#1}}

\def\ms{\hspace{0.05em}\mathrm{ms}}
\def\s{\hspace{0.05em}\mathrm{s}}
\def\p{\hspace{0.05em}\%}
\def\K{\hspace{0.1em}\mathrm{K}}
\def\M{\hspace{0.1em}\mathrm{M}}
\def\B{\hspace{0.1em}\mathrm{B}}
\def\GB{\hspace{0.1em}\mathrm{GB}}

\def\QLever{QLever}
\def\Virtuoso{Virtuoso}
\def\Blazegraph{Blazegraph}

\def\entities{{\bf \%entities\%}}
\def\predicates{{\bf \%predicates\%}}
\def\prefix{{\bf \%prefix\%}}
\def\context{{\bf \%context\%}}

\def\subject{{\bf \%subject\%}}
\def\predicate{{\bf \%predicate\%}}

\setcopyright{rightsretained}

\begin{document}

\title{Efficient SPARQL Autocompletion via SPARQL}

\author{Hannah Bast, Johannes Kalmbach, Theresa Klumpp, Florian Kramer, Niklas Schnelle}

\affiliation{%
  \institution{University of Freiburg}
  \city{Freiburg}
  \country{Germany}
}
\email{{bast, kalmbacj, klumppt, kramerfl, schnelle}@cs.uni-freiburg.de}

\renewcommand{\shortauthors}{H. Bast et al.}

\begin{abstract}
We show how to achieve fast autocompletion for SPARQL queries on very large knowledge bases.
At any position in the body of a SPARQL query, the autocompletion suggests matching subjects, predicates, or objects.
The suggestions are context-sensitive in the sense that they lead to a non-empty result and are ranked by their relevance to the part of the query already typed.
The suggestions can be narrowed down by prefix search on the names and aliases of the desired subject, predicate, or object.
All suggestions are themselves obtained via SPARQL queries, which we call \emph{autocompletion queries}.
For existing SPARQL engines, these queries are impractically slow on large knowledge bases.
We present various algorithmic and engineering improvements of an existing SPARQL engine
such that these autocompletion queries are executed efficiently.
We provide an extensive evaluation of a variety of suggestion methods on three large knowledge bases,
including Wikidata (6.9B triples).
We explore the trade-off between the relevance of the suggestions and the processing time of the autocompletion queries.
We compare our results with two widely used SPARQL engines, Virtuoso and Blazegraph.
On Wikidata, we achieve fully sensitive suggestions with sub-second response times for over $90\%$ of a large and diverse set of thousands of autocompletion queries.
Materials for full reproducibility, an interactive evaluation web app, and a demo 
are available on: \url{http://ad.informatik.uni-freiburg.de/publications}.
\end{abstract}

\settopmatter{printfolios=true}

\maketitle

\section{Introduction}\label{sec:introduction}

\begin{figure*}
\includegraphics[width=0.45\textwidth]{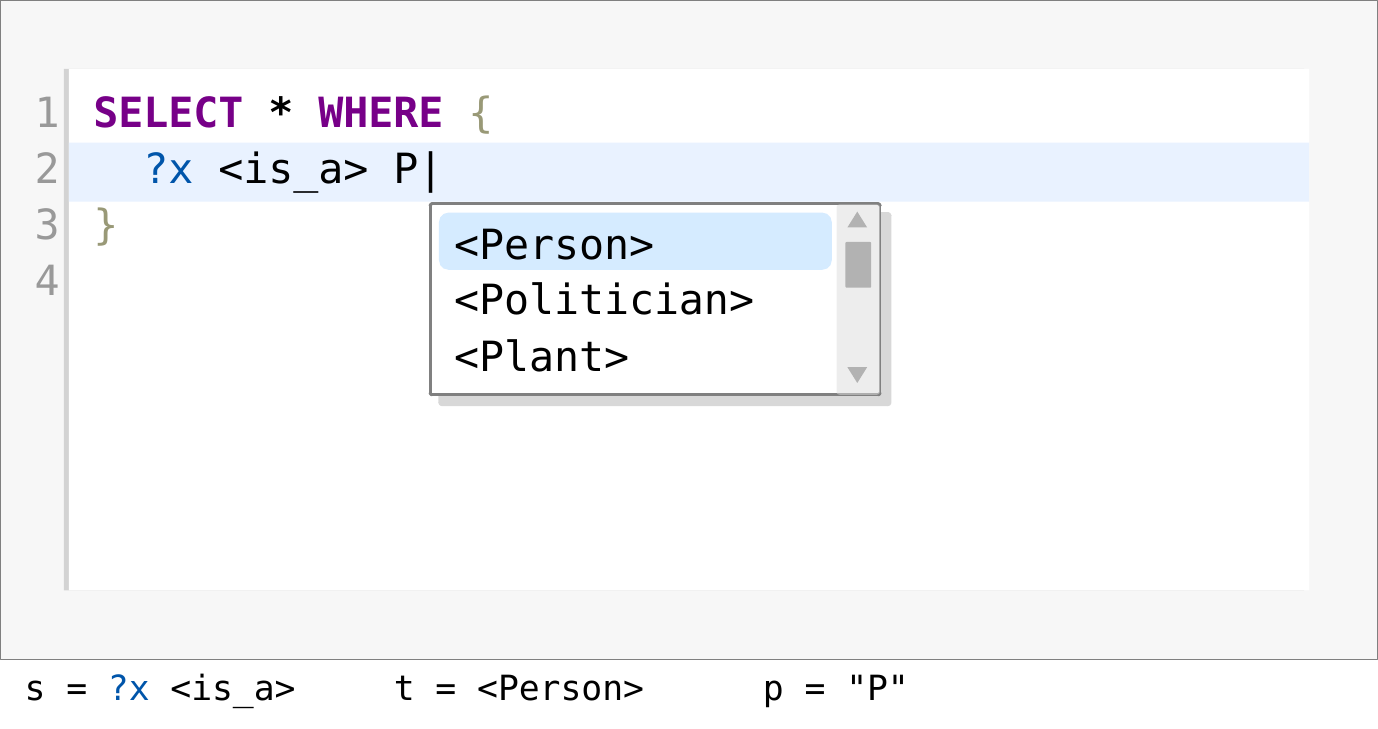}
\hspace{0.02\textwidth}
\includegraphics[width=0.45\textwidth]{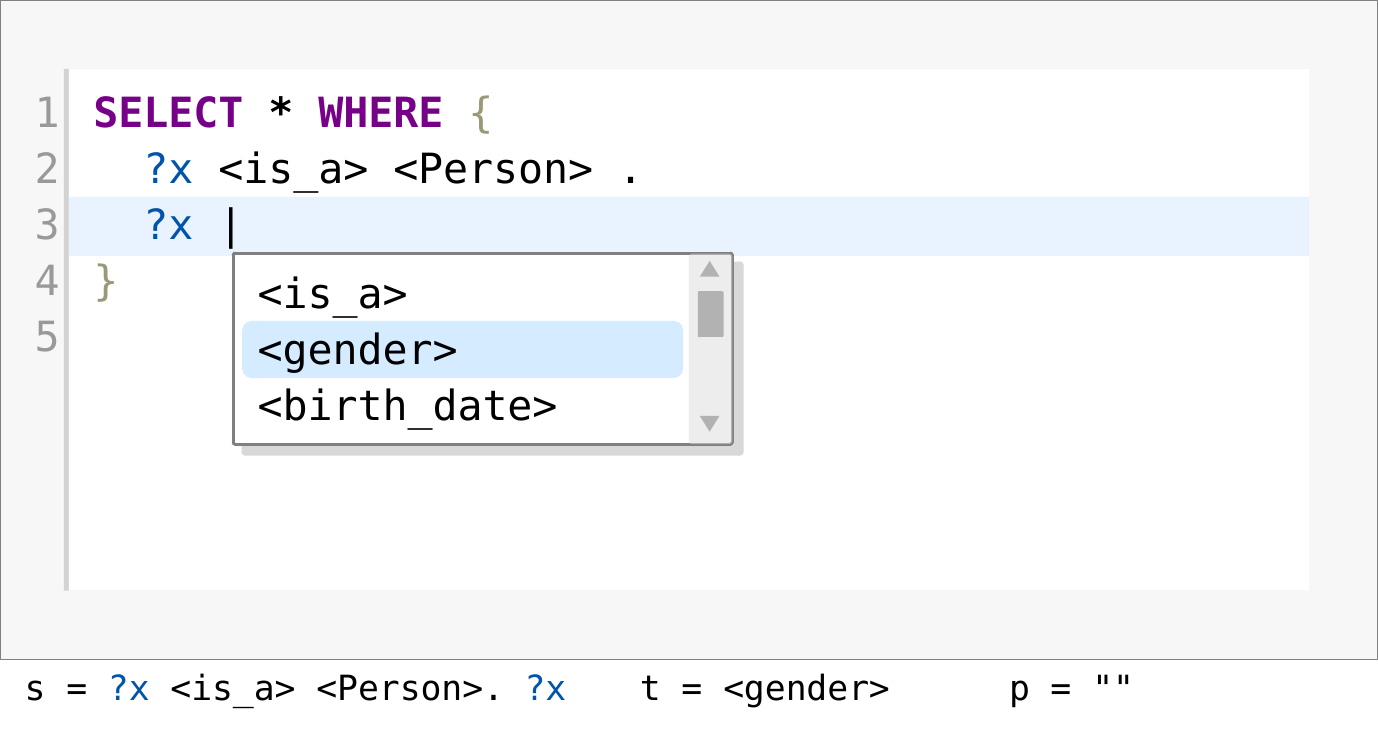}\\
\vspace{0.02\textwidth}
\includegraphics[width=0.45\textwidth]{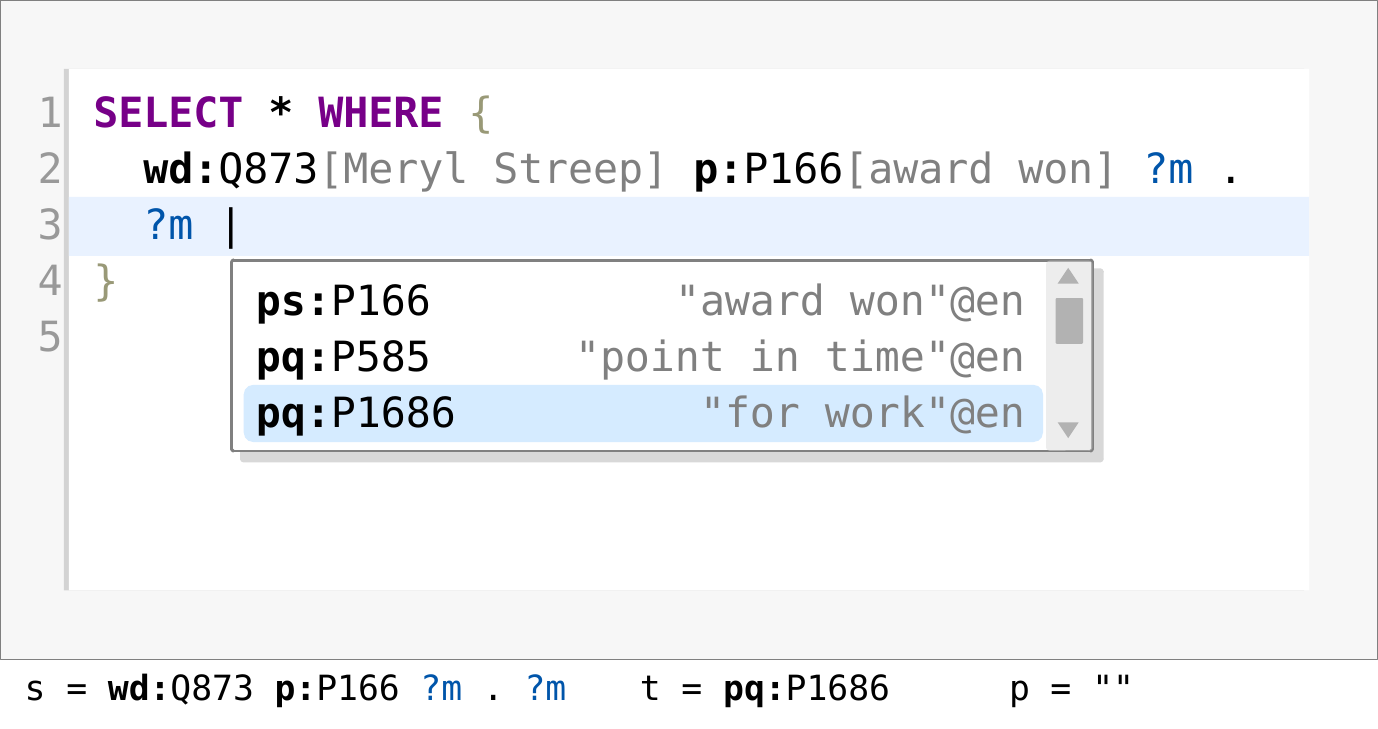}
\hspace{0.02\textwidth}
\includegraphics[width=0.45\textwidth]{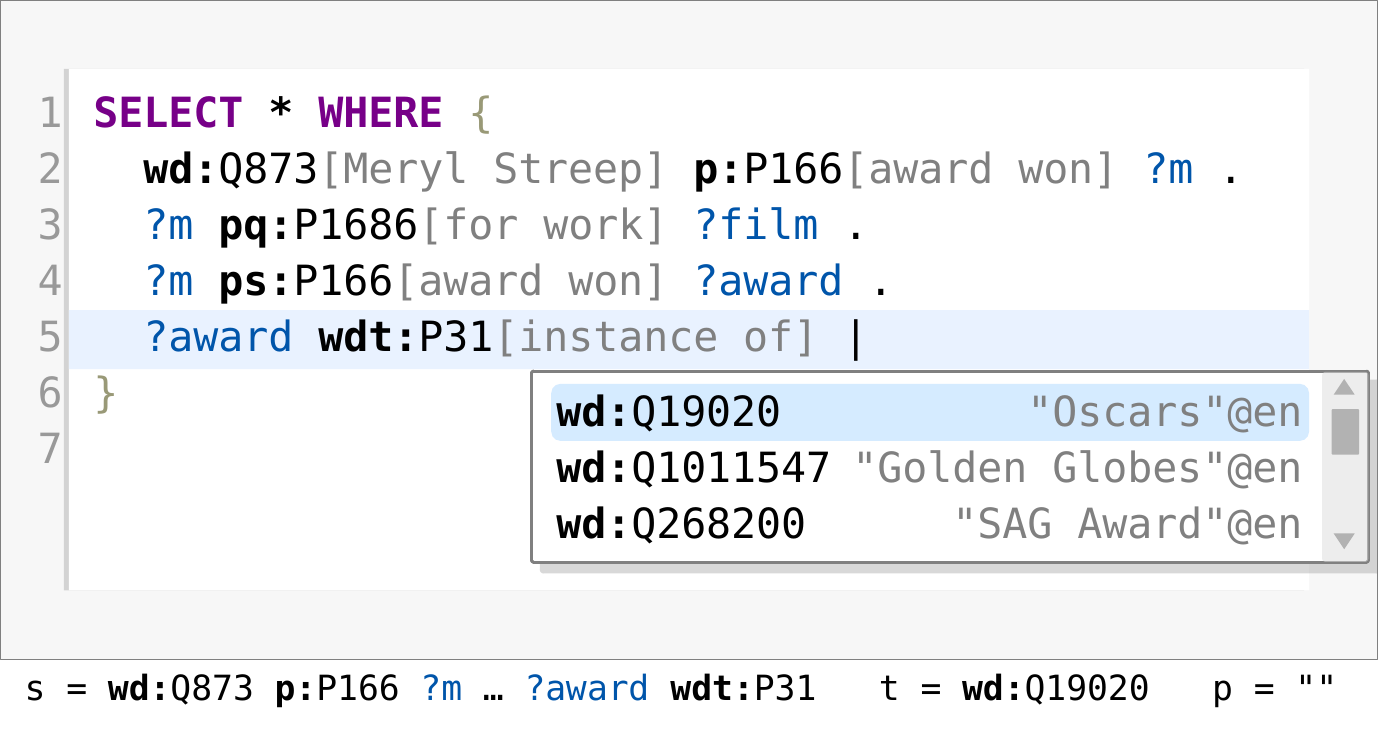}
\vspace{-2mm}
\caption{Four screenshots of our autocompletion in action, with three suggestions each.
Top-left and top-right: Examples 1 and 2 from Section \ref{sec:definition}, for a query on Fbeasy,
where IRIs are understandable for a human.
Bottom-left and bottom-right: Example 3 from Section \ref{sec:definition} and a continuation not described in the text,
for a query on Wikidata, where IRIs are alpha-numeric.
The assignments below each screenshot show the values of the variables from our problem definition:
$s$ (the part of the SPARQL query already typed), $t$ (the desired token), and $p$ (a prefix of a name or alias of that token).
For a live demo, see \url{https://qlever.cs.uni-freiburg.de}.}
\label{fig:qlever-ui}
\vspace{-1mm}
\end{figure*}

Knowledge bases play an increasingly important role in modern retrieval 
systems.
The prevailing data model is the Resource Description Framework (RDF),
where the data is stored as subject-predicate-object triples.
Each subject, predicate, or object is either an
\emph{Internationalized Resource Identifier} (IRI), enclosed in angle brackets,
or a so-called \emph{literal}, enclosed in quotes.
For example:
%
\begin{sparqlkb}
$<$Meryl\_Streep$>$          \> $<$is\_a$>$        \> $<$Person$>$ \\
$<$Meryl\_Streep$>$          \> $<$gender$>$       \> $<$Female$>$ \\
$<$Meryl\_Streep$>$          \> $<$award\_won$>$   \> $<$Oscar\_Best\_Actress$>$ \\
$<$Meryl\_Streep$>$          \> $<$birth\_date$>$  \> "1949-06-22" \\
$<$Ang\_Lee$>$               \> $<$award\_won$>$   \> $<$Oscar\_Best\_Director$>$ \\
$<$Oscar\_Best\_Actress$>$   \> $<$is\_a$>$        \> $<$Oscar$>$ \\ 
$<$Oscar\_Best\_Director$>$  \> $<$is\_a$>$        \> $<$Oscar$>$
\end{sparqlkb}
%
RDF data allow queries with precise semantics.
For example, the following query finds all women who won an Oscar.
The query is formulated in SPARQL, the standard query language for RDF data.%
\begin{sparqlq}
{\SELECT} ?entity ?award {\WHERE} \{ \\
\> ?entity  \> $<$is\_a$>$  \>  <Person> \> . \\
\> ?entity  \> $<$gender$>$       \>  <Female> \> . \\
\> ?entity  \> $<$award\_won$>$    \>  ?award   \> . \\
\> ?award   \> $<$is\_a$>$  \>  <Oscar>  \> \}
\end{sparqlq}
The result for this query is a table with two columns, where each row contains the name of the person and the name of the award.
For the tiny example knowledge base above, the result is:
\begin{sparqlkb}
$<$Meryl\_Streep$>$ $<$Oscar\_Best\_Actress$>$
\end{sparqlkb}
The seven example triples above come from \emph{Fbeasy} ($362\M$ triples) \cite{DBLP:conf/www/BastBBH14},
an easy-to-use subset of \emph{Freebase} ($1.9\B$ triples) \cite{DBLP:conf/sigmod/BollackerEPST08}.
The largest general-purpose knowledge base to date is \emph{Wikidata} ($6.9\B$ triples, as of 07-01-2020)
\cite{DBLP:journals/cacm/VrandecicK14}.
Fbeasy has human-readable IRIs for all entities.
In Wikidata, almost all IRIs are alpha-numeric and human-readable names can be obtained via dedicated predicates;
see the example query below.
Freebase uses a mix of human-readable and alpha-numeric IRIs.
We consider all three knowledge bases in this paper; see Section \ref{sec:evaluation:datasets} for details.

SPARQL is conceptually easy,
because queries can be formulated as lists of triples, just like the data.
However, finding the IRIs relevant for a query is often hard in practice.
This is true even for human-readable IRIs when there are very many of them.
It becomes extremely hard when IRIs are alpha-numeric.
For example, consider the seemingly simple request: the Oscars of Meryl Streep and the movies she won them for.
The correct SPARQL query on Wikidata is quite complex (an explanation follows below the query).
\def\bb#1{{\color{grey} [#1]}}
\begin{sparqlq}
{\SELECT} ?award ?film {\WHERE} \{ \\
\> wd:Q873 p:P166 ?m . \\
\> ?m pq:P1686 ?film\_id . \\
\> ?m ps:P166 ?award\_id . \\
\> ?award\_id wdt:P31 wd:Q19020 . \\
\> ?award\_id rdfs:label ?award . \\ 
\> ?film\_id rdfs:label ?film \} 
\end{sparqlq}
The \emph{wd:}, \emph{wdt:}, etc. are IRI prefixes. We have omitted their definition here to save space; see \url{https://en.wikibooks.org/wiki/SPARQL/Prefixes}.
The IRIs \emph{wd:Q873} and \emph{wd:Q19020} denote \emph{Meryl Streep} and the \emph{Academy Awards}, respectively.
The IRIs \emph{p:P166} and \emph{ps:P166} both represent the predicate \emph{award received}.
In Wikidata, a predicate with prefix \emph{p:} connects a subject to a statement node,
which can then be connected to several objects, thus representing an $n$-ary relation between the subject and the objects.
The predicate with prefix \emph{ps:} connects the subject to the ``main object'' of the relation; the IRI of the award in this case.
Predicates with prefix \emph{pq:} connect the subject to so-called ``qualifying'' information; the IRI of the film in this case (there is also a \emph{pq:} predicate for the point in time).
The IRI \emph{wdt:P31} stands for the \emph{instance of} relation.
The \emph{rdfs:label} predicate connects entities to their human-readable names.




\subsection{Problem Definition and Three Examples}\label{sec:definition}

The goal of this paper is to assist the user in typing the body of a SPARQL query
by providing suggestions at any point in the query.
The suggestions should be ranked by relevance to the part of the query already typed.
We first provide a formal problem definition and then explain it at length via three examples.

\def\sparql{s} 
\def\token{t} 
\def\prefixvar{p} 

\medskip\noindent{\bf Definition.}
Imagine a valid SPARQL query typed until a certain point, called the \emph{cursor position}.
Let $\sparql$ be the part of the body of the \underline{S}PARQL query until that point.
Let $\token$ be the complete \underline{t}oken (subject, predicate, or object) at the cursor position.
Let $\prefixvar$ be a \underline{p}refix of a name or alias of $t$, possibly empty.
The \emph{SPARQL Autocompletion via SPARQL} problem is:
Given $\sparql$ and $\prefixvar$, construct and process a SPARQL query,
called \emph{autocompletion query} or \emph{AC query}, with the following properties:\\[2mm]
1. The AC query returns a table with each row corresponding to a suggestion and the following three columns:\\[1mm]
\emph{?entity} (an entity from the knowledge base),\\
\emph{?name} (a name or alias of that entity, starting with $\prefixvar$),\\
\emph{?score} (an estimate of how likely that entity is at the cursor position).\\[1mm]
2. The rows of the table are sorted in descending order of \emph{?score}.\\[1mm]
3. One of the rows contains $\token$ in the \emph{?entity} column.\\[2mm]
There are two objectives:\\[.5mm]
\emph{Relevance}: Each suggestion should be \emph{context-sensitive}%
\footnote{In the following, we often just call such suggestions \emph{sensitive}.}
in the sense that it continues the SPARQL query in a meaningful way, that is,
such that there exists a continuation with a non-empty result.
The row with the desired $\token$ should be as high up in the table as possible.\\[1mm]
\emph{Efficiency}: The query should be processed as quickly as possible.

\vspace{2.5mm}\noindent
We next illustrate this definition by three examples, which are also depicted in Figure \ref{fig:qlever-ui}.
We also discuss running times for various SPARQL engines.
These engines are described in Section \ref{sec:efficiency:engines},
and our experimental setup is described in Section \ref{sec:evaluation:setup}.


\medskip\noindent
{\bf Example 1}
Assume we have typed the body of the first SPARQL query from the introduction until before the first object; see below.
This is the $s$ from the definition.
The $\cursor$ symbol marks the cursor position and the prefix $\prefixvar$ is \emph{"P"}.
The token $\token$ we are looking for at this position is \emph{$<$Person$>$}.
The knowledge base is Fbeasy. 
%
\begin{sparqlq}
\> ?x $<$is\_a$>$ P\cursor
\end{sparqlq}
The following AC query computes a table containing each object \emph{?entity} and its name \emph{?name},
such that the name starts with \emph{P}
and the triple \emph{?x $<$is\_a$>$ ?entity} exists.
The table is sorted in descending order of the number of such triples
for each \emph{?entity}.
\begin{sparqlq}
{\SELECT} ?entity ?name (\COUNT(?x) {\AS} ?score) {\WHERE} \{\\
\> ?x $<$is\_a$>$ ?entity . \\
\> BIND (STR(?entity) AS ?name) . \\
\> {\FILTER} {\REGEX}(?name, "\verb|^|P")  \\
\} {\GROUPBY} ?entity ?name {\ORDERBY} DESC(?score)
\end{sparqlq}
The first three result rows for that query look as follows.
Note that for this knowledge base, the name of an entity is simply the IRI, interpreted as a string
(that is what the \emph{STR} function does).
\begin{sparqlkb2}
$<$Person$>$     \> "Person" 			\>  "3970825" \\
$<$Politician$>$ \> "Politician"  \>  "127809"  \\
$<$Plant$>$      \> "Plant"  			\>  "60459"
\end{sparqlkb2}
Let us discuss the objectives from our definition for this AC query.\\[1mm]
\emph{Relevance:} Perfect, because the desired token $t$ appears in the first row of the table,
and, by construction, all suggested entities lead to a non-empty result.\\[1mm]
\emph{Efficiency:} Most query engines, including our competitors Virtuoso and Blazegraph,
process this query as follows:
First materialize a table \emph{?x ?entity ?name} with all subjects and objects of the \emph{$<$is\_a$>$} predicate
and the name of these objects.
On Fbeasy, this table has 130M rows. 
Then check for each row whether the name matches the regular expression.
This is slow: it takes $20\s$ on Virtuoso and $48\s$ on Blazegraph.
Our engine works with internal IDs for the strings, which are pre-ordered in lexicographical order of the strings.
That way, we can maximally delay a materialization of the strings and perform prefix matches via binary search.
This is a major performance factor, not only for AC queries.
In particular, the AC query above takes only $0.3\s$.
More details in Section \ref{sec:efficiency}.



\medskip\noindent
{\bf Example 2}
Now assume that we have typed the SPARQL query a little bit further.
The following is now our $\sparql$ and the desired token $\token$ at the cursor position is \emph{$<$gender$>$}.
The prefix $\prefixvar$ is empty.
\begin{sparqlq}
\> ?x $<$is\_a$>$ $<$Person$>$ . \\
\> ?x \cursor
\end{sparqlq}
The following AC query gives us a ranked list of \emph{predicates} that lead to a non-empty result.
The score for each predicate is the number of persons (that is, entities matching the first triple) that have a triple with that predicate.\footnote{If a person has several triples with the same predicate, we only count the predicate once, hence the {\DISTINCT}. This is explained in more detail in Section \ref{sec:ac_queries}.}
\begin{sparqlq}
{\SELECT} ?entity ?name ({\COUNT}({\DISTINCT} ?x) {\AS} ?score) {\WHERE} \{\\
\> ?x $<$is\_a$>$ $<$Person$>$ . \\
\> ?x ?entity []~.       								\\
\> BIND (STR(?entity) AS ?name) . \\
\} {\GROUPBY} ?entity ?name {\ORDERBY} DESC(?score)
\end{sparqlq}
The first three result rows for this AC query are as follows:
\begin{sparqlkb2}
$<$is\_a$>$ 	 			 \>  "is\_a" 					 \>  "3970825"  \\
$<$gender$>$  			 \>  "gender"				   \>  "2276146"   \\
$<$birth\_date$>$    \>  "birth\_date"     \>  "1915167" 
\end{sparqlkb2}
\emph{Relevance:} The desired token $\token$ is second in this table
and, again by construction, the AC query only returns predicates that lead to a hit.
The suggestions are ranked by how often each of them occurs with the set of entities
defined by the part of the query already typed.
Because of this, we did not have to type a single letter here to get very good suggestions.
In Section \ref{sec:evaluation}, we will also evaluate so-called \emph{agnostic} AC queries.
An agnostic AC query computes a list of \emph{all} predicates, ordered by frequency,
but independent of the rest of the query.
Such a query would have bad relevance here:
\emph{$<$gender$>$} would not be among the top suggestions,
and there would be many suggestions that are not meaningful for a person,
for example \emph{$<$release$>$} (of a musical recording).\\[1mm]
%
\emph{Efficiency:} Most query engines, including our competitors Virtuoso and Blazegraph,
process this AC query as follows.
The two triples from the query body are joined into a large table,
with one row for each triple of each person.
In Fbeasy this table has 37M rows.
Producing a table of that size is slow and so is determining the predicate counts from that table.
Without the {\BIND} and the {\DISTINCT}, Virtuoso takes $2.4\s$ and Blazegraph takes $35\s$.%
\footnote{Neither Virtuoso nor Blazegraph can handle the query with the {\DISTINCT},
apparently because that would require a full sort of the table.
However, suggestions are also reasonable (though not quite as good) without the {\DISTINCT}.}
%
%
%
%
%
For the corresponding AC query on Wikidata, the table has $196\M$ rows.  
Virtuoso can process it in $5.8\s$, while Blazegraph aborts with an out-of-memory error after $85\s$.
%
%
%
%
%
Our query engine avoids iterating over all the triples for each person.
Instead, our engine makes use of a query-independent pre-processing
that identifies groups of entities with the exact same set of predicates (called a \emph{pattern}).
This trick is described in Section \ref{sec:patterns}.
It enables a query time of below $0.1\s$ for the AC query above on Fbeasy,
and $0.6\s$ for its counterpart on Wikidata.
%
%
%
%
%
%
%


%

\medskip\noindent
{\bf Example 3}
Our last example is based on Wikidata, where entities have alpha-numeric IRIs,
and names and aliases are obtained via dedicated predicates \emph{rdfs:name} and \emph{skos:altLabel}.
Assume that we have typed the body of our second SPARQL query from the introduction this far:
\begin{sparqlq}
\> wd:Q873 p:P166 ?m . \\
\> ?m \cursor
\end{sparqlq}%
This is our $\sparql$ and the prefix $\prefixvar$ is again empty.
Recall that \emph{wd:Q873} stands for \emph{Meryl Streep} and \emph{p:P166} connects this entity to
all statement nodes \emph{?m} pertaining to one of her awards.
The desired token $\token$ is \emph{pq:P1686},
which leads us to the awarded films. 
The following AC query gives us a list of predicates (and their names) that lead to results at this point.
The score is analogous to that of the previous example.
The predicate path {\it \verb|^|(<>|!<>)/rdfs:label} gives us the label of a predicate in Wikidata;
see Section \ref{sec:ac_queries:building_blocks}.
\begin{sparqlq}
{\SELECT} ?entity ?name ({\COUNT}({\DISTINCT} ?m) {\AS} ?score) {\WHERE} \{\\
\> wd:Q873 p:P166 ?m .  \\
\> ?m ?entity [] .   		\\
\> ?entity \^{}(<>|!<>)/rdfs:label ?name \\
\} {\GROUPBY} ?entity ?name {\ORDERBY} DESC(?score)
\end{sparqlq}
The first three result rows for this AC query look as follows:
\begin{sparqlq}
ps:P166 	\>\>  "award received" 	\> "33"  \\
pq:P585  	\>\>  "point in time"  	\> "23"  \\
pq:P1686  \>\>  "for work" 				\> "10"  
\end{sparqlq}
These rows tell us that Wikidata knows about $33$ awards of Meryl Streep,
the point in time for $23$ of them, 
and for which work the award was given for $10$ of them.

\emph{Relevance:} Again perfect for this AC query.
And note how helpful the suggestions are!
Without these suggestions it would require extremely intimate knowledge of Wikidata to know that
we need the predicate suffixes \emph{P166} and \emph{P1686}
and the prefixes \emph{ps:} (which stands for the main entity of a statement node)
and \emph{pq:} (which stands for additional properties of a statement node).
Agnostic suggestions, as briefly mentioned in Example 2
and defined in Section \ref{sec:ac_queries:agnostic}, would not be of much help here.

\emph{Efficiency:} The query is easy because there are only few bindings for \emph{?m},
namely one for each of Meryl Streep's awards.
With optimal query planning\footnote{When equivalently rewriting the query, such that the result from the first two triples are aggregated in a subquery (which corresponds to the best query plan), Blazegraph inexplicably takes forever.}
and the \emph{rdfs:label} triples in main memory,
Blazegraph processes the query in around $100\ms$, Virtuoso takes $10\ms$,
and our engine takes $1\ms$.

%

%

%

\def\calS{\mathcal{S}}
\def\calP{\mathcal{P}}
\def\prefixes{\Sigma}

\subsection{Scope and limits of our definition}\label{sec:introduction:scope}

Our AC queries provide suggestions for the IRIs and literals in a SPARQL query.
A typical user interface for SPARQL autocompletion also involves suggestions
for variable names or SPARQL constructs like {\OPTIONAL}, {\FILTER}, {\UNION} or {\GROUPBY}
at appropriate positions in the query.
These suggestions are not particularly challenging with respect to relevance or efficiency.
However, our AC queries do have to deal properly with these constructs, when they occur in the query;
see Section \ref{sec:ac_queries:extend}.

Our definition assumes that queries are typed from top to bottom
and that each triple is entered from left to right (subject, predicate, object).
It would be straightforward to drop those constraints using variants of our AC queries.
However, it's not strictly necessary for the incremental construction of a SPARQL query
and it would significantly complicate our definition and explanations in this paper.
We therefore omitted this use case from our definition.



\subsection{Our contributions}\label{sec:contributions}

We consider the following as our main contributions:

\vspace{0mm}
\begin{itemize}[parsep=1.0mm,leftmargin=0mm,itemindent=4mm]

\item We develop the idea of providing SPARQL autocompletion via SPARQL queries.
These \emph{AC queries} can be processed by any (standard-conforming) SPARQL engine.
The basic idea is already found in previous work, but in less generality.
See Sections \ref{sec:related} and \ref{sec:ac_queries}.

\item We extend an existing query engine, called \emph{\QLever}, such that these AC queries can be processed efficiently.
The extension is technically challenging and comprises both algorithmic ideas and algorithm engineering.
See Section \ref{sec:efficiency}.

\item We provide an extensive evaluation on three large knowledge bases, including Wikidata (6.9B triples).
We explore a variety of AC queries in depth,
in particular, their relevance and efficiency and the trade-off between these two objectives.
We compare our results with two widely used SPARQL engines, Virtuoso and Blazegraph,
which we outperform by a large margin.
On Wikidata, we achieve fully sensitive suggestions with sub-second response times
for over 90\% of a large and diverse set of thousands of autocompletion queries.
See Section \ref{sec:evaluation}.



\item We provide materials for full reproducibility:
code, queries, indexes, result files, and a web application that allows an interactive exploration
of all the details of our experimental results.
In particular, the web application provides the {\LaTeX} code for Table \ref{tab:main_results} with a single click.
These materials are available under \url{https://ad.informatik.uni-freiburg.de/publications}.

%


\end{itemize}


\input{related-work}

\section{AC Query Templates}\label{sec:ac_queries}

In this section, we describe the templates of the various AC queries
mentioned in Section \ref{sec:introduction} and used in our evaluation in Section \ref{sec:evaluation}.
In Section \ref{sec:evaluation:queries}, we will derive actual AC queries from these templates.

In Section \ref{sec:ac_queries:building_blocks}, 
we first specify two easy-to-understand queries {\entities} and {\predicates}.
In Sections \ref{sec:ac_queries:sensitive} - \ref{sec:ac_queries:mixed},
we will use them as building blocks to construct our actual AC query templates.
For the sake of explanation, we first assume SPARQL queries, where the body consists only of triples.
In Section \ref{sec:ac_queries:extend}, we then explain how to extend this to more complex SPARQL queries.


\subsection{Building blocks for Wikidata}\label{sec:ac_queries:building_blocks}

The building block {\entities} computes for each entity its name and aliases and a score.
Here is the query for Wikidata:
%
\begin{sparqlqq}
1. \> {\SELECT} ?entity ?name ({\COUNT}(?x) {\AS} ?score) {\WHERE} \{ \\
2. \>\> ?entity rdfs:label|skos:altLabel ?name . \\
3. \>\> ?entity \verb|^|schema:about ?x . \\
4. \> \} {\GROUPBY} ?entity ?name
\end{sparqlqq}
For the score, we use the number of Wikipedia articles about an entity (line 3).
For Freebase, the predicates in lines 2 and 3 are replaced by \emph{fb:type}\emph{.object.name|fb:common.topic.alias} 
and \emph{fb:type.object.type}, respectively.
For Fbeasy, line 2 is replaced by \emph{{\BIND}({\STR}(?entity) {\AS} ?name)}
and the predicate in line 3 is replaced by \emph{$<$is\_a$>$}.
That is, for Fbeasy and Freebase we use the number of types as score.
These scores are only needed for subject AC queries (see Section \ref{sec:ac_queries:sensitive})
and agnostic AC queries (see Section \ref{sec:ac_queries:agnostic}).

The building block {\predicates} computes a name and a score for each predicate.
Here is the query for Wikidata:
\begin{sparqlqq}
1. \> {\SELECT} ?entity ?name ?score {\WHERE} \{ \\
2. \>\> \{ {\SELECT} ?entity ({\COUNT}({\DISTINCT} ?x) {\AS} ?score) {\WHERE} \{ \\
3. \>\>\>  ?x ?entity [] \\
4. \>\> \} {\GROUPBY} ?entity \} \\
5. \>\> ?entity (\verb|^|(<>|!<>)/(rdfs:label|skos:altLabel))? ?name \\
6. \>\} 
\end{sparqlqq}
As score for each predicate, we take the number of distinct subjects of that predicate.
The name of a predicate is obtained via a special (predicate-dependent) property.
From the subject of that property, we then obtain the name via the usual predicates.
The {\it \verb|^|(<>|!<>)} in line 5 is just a SPARQL shorthand of saying:
follow any predicate in reverse direction.
Some predicates, like \emph{rdf:type} have no explicit name at all,
in which case we take the IRI as the name,
hence the $(...)?$ around the predicate path.
For Freebase, we replace line 5 by \emph{?entity fb:type.object.name ?name}.
For Fbeasy, all IRIs are human-readable and we replace line 5 by \emph{{\BIND}(STR(?entity) AS ?name)}.


\def\calS{{\mathcal S}}
\subsection{Sensitive AC queries}\label{sec:ac_queries:sensitive}

To construct an AC query, we can make use of the part $s$ of the SPARQL query body already typed
(see our definition in Section \ref{sec:definition}).
It is important that we only use the part of $s$
that is actually ``connected'' to the triple at the current cursor position.%
\footnote{In SPARQL, the result of two disconnected graph patterns is the cross product,
which can be become huge very easily and is usually not what one wants.}
We call this part the {\context} and next explain how it is computed.

Let $T$ be the partial triple that is currently being typed.
Let $\calS$ be the set of all finished triples and FILTER clauses inside $s$.
Note that $\calS$ also includes triples and filters that appear inside SPARQL constructs
like {\OPTIONAL}, {\UNION}, {\MINUS}, and sub-queries.
Construct an undirected graph with node set ${\calS} \cup \{T\}$
and an edge between two nodes if they share a variable.%
\footnote{We also take care of scoping and variable renaming introduced by subqueries.}
Then {\context} is $s$ without $T$ and without all nodes in ${\calS}$ that are not reachable from $T$.
It can be easily computed with, for example, a breadth-first search starting from $T$.


\def\scoreT{{\bf \%score\%}}
\def\xT{{\bf \%x\%}}

Here is the AC query template for a subject with a non-empty {\prefix}
(the $p$ from our definition).
This template is used rarely because in typical SPARQL queries,
most triples have a variable as subject.
%
\begin{sparqlqq}
1. \> {\SELECT} ?entity ({\SAMPLE}(?name) {\AS} ?name) \\
2. \> \phantom{{\SELECT} ?entity} ({\SAMPLE}(?score) {\AS} ?score) {\WHERE} \{ \\
3. \>\> \{ {\entities} \} {\FILTER} {\REGEX}(?name, \verb|"|\verb|^|{\prefix}\verb|"|) \\
4. \> \} {\GROUPBY} ?entity {\ORDERBY} {\DESC} (?score)
\end{sparqlqq}

\smallskip\noindent
Here is the AC query template for a predicate for given {\prefix}, {\context}, 
and {\subject} (the subject of the triple at the current cursor position).
If {\context} is empty and {\subject} is a variable, lines 3-5 are (redundant and) omitted and \emph{?score\_2} in line 2 is replaced by \emph{?score}.
If {\subject} is a variable, {\xT} is \emph{{\DISTINCT} {\subject}}, otherwise \emph{?object}.
This template generalizes Examples 2 and 3 from Section \ref{sec:definition},
which also provide an intuition for the score.
\begin{sparqlqq}
1. \> {\SELECT} ?entity ({\SAMPLE}(?name) {\AS} ?name) \\
2. \> \phantom{{\SELECT} ?entity} ({\SAMPLE}(?score\_2) {\AS} ?score) {\WHERE} \{ \\
3. \>\> \{ {\SELECT} ?entity ({\COUNT}({\xT}) {\AS} ?score\_2) {\WHERE} \{ \\
4. \>\>\>  {\context} {\subject} ?entity ?object \\
5. \>\> \} {\GROUPBY} ?entity \} \\
6. \>\> \{ {\predicates} \} {\FILTER} {\REGEX}(?name, \verb|"|\verb|^|{\prefix}\verb|"|) \\
7. \> \} {\GROUPBY} ?entity {\ORDERBY} {\DESC} (?score)
\end{sparqlqq}
%
%
%

\smallskip\noindent
Here is the AC query template for an object for given {\prefix}, {\context},
{\subject}, and {\predicate} (the predicate of the triple at the current cursor position).
This template generalizes Example 1 from Section \ref{sec:definition},
which also provides an intuition for the score.
%
\begin{sparqlqq}
1. \> {\SELECT} ?entity ({\SAMPLE}(?name) {\AS} ?name) \\
2. \> \phantom{{\SELECT} ?entity} ({\SAMPLE}(?score\_2) {\AS} ?score) {\WHERE} \{ \\
3. \>\> \{ {\SELECT} ?entity ({\COUNT}(?entity) {\AS} ?score\_2) {\WHERE} \{ \\
4. \>\>\>  {\context} {\subject} {\predicate} ?entity  \\
5. \>\> \} {\GROUPBY} ?entity \} \\
6. \>\> \{ {\entities} \} {\FILTER} {\REGEX}(?name, \verb|"|\verb|^|{\prefix}\verb|"|) \\
7. \> \} {\GROUPBY} ?entity {\ORDERBY} {\DESC} (?score)
\end{sparqlqq}
%
%
%

%
%

\subsection{Agnostic and Unranked AC queries}\label{sec:ac_queries:agnostic}

We first define the agnostic AC queries.
For a subject and a predicate, the agnostic AC query is like the respective sensitive queries, but with empty {\context}.
The agnostic AC query for an object is identical to the agnostic AC query for a subject.

The result for an agnostic AC query is then exactly the result of the precomputed {\entities} or {\predicates},
filtered by {\prefix}.
In Section \ref{sec:evaluation}, we will see that {\QLever} can always process these queries in time below one second.
Note that some work is required at query time if {\prefix} is non-empty:
we then need to filter out the results matching the prefix (from a table sorted by entity name),
which can be many if the prefix is short or common.
Also note that, in principle, we could compute the results for these queries
with a special-purpose data structure, outside of the SPARQL engine.%
\footnote{In fact, Blazegraph uses such a special-purpose data structure
to compute names for entities to avoid the expensive join with the huge \emph{rdfs:label} predicate.}
However, it is convenient if we don't need such a data structure and can just use the engine
that we have in place anyway.

The unranked AC queries are just like the agnostic AC queries, but without the final {\ORDERBY}.
They have no practical relevance, but we include them in our evaluation in Section \ref{sec:evaluation} as a baseline,
to show how important ranking of suggestions 
(neglected in several previous works for efficiency reasons, see Section \ref{sec:related}) is.

\subsection{Mixed AC queries}\label{sec:ac_queries:mixed}

The suggestions of sensitive AC queries are clearly preferable,
but as we will see in our evaluation in Section \ref{sec:evaluation},
some of them are very hard to compute, especially on a large knowledge base like Wikidata.
In contrast, agnostic AC queries are always fast, at the price of a much lower relevance.

We therefore also consider a \emph{mixed} mode, where we simultaneously issue an agnostic and a sensitive query.
If the sensitive query finishes within a given timeout, we take that result.
Otherwise, we take the result from the agnostic query.
We pick a timeout of $1\s$, since users are rarely willing to wait longer for an interactive suggestion.
Since we can answer all agnostic queries in time $1\s$, we then always get suggestions within $1\s$,
though not always sensitive.
Mixed AC queries are thus a compromise between relevance and efficiency.
The faster the engine, the better the compromise.
More about this in Section \ref{sec:evaluation}.


\subsection{AC queries for complex SPARQL constructs}\label{sec:ac_queries:extend}

In our definition of {\context} in Section \ref{sec:ac_queries:sensitive},
we have assumed that the triple $T$ at the current cursor position is not inside
of an {\OPTIONAL}, {\UNION}, {\MINUS}, or a sub-query and does not contain a property path. 
We handle these cases as follows.\\[1mm]
1. If $T$ is inside of an {\OPTIONAL} clause, we treat it as if it were outside of the {\OPTIONAL} clause.
This means that we only get suggestions that lead to at least one result (without NULL values).\\[.5mm]
2. If $T$ is inside a {\UNION} clause, we also treat it as if it were outside of the clause.
Additionally, triples from the other ``branch'' of the same {\UNION} are excluded from the context.
That way we ensure that each branch contributes at least one result to the query.\\[.5mm]
3. If $T$ is inside a {\MINUS} clause, we ignore the {\MINUS}
and give suggestions as if the triple was added and not subtracted.
As a consequence, we rank those suggestions highest,
that would \emph{remove} the highest number of results when added to the query using {\MINUS}.\\[.5mm]
4. If $T$ is inside a sub-query (SPARQL queries nested inside the body of another SPARQL query),
the surrounding query is ignored.
We could also incorporate the surrounding query into the {\context},
but found this to have negligible benefit in our evaluation.\\[.5mm]
5. A triple with a property path like \emph{wdt:P31/wdt:P279*}
(\emph{instance of}, followed by any number of \emph{subclass of})
is equivalently transformed into two triples with a single predicate each.
We then launch an AC query for each of these predicates (one after the other).\\[1mm]
Note that by handling {\OPTIONAL}, {\UNION} and {\MINUS} this way,
we achieve more than the relevance objective from our definition in Section \ref{sec:definition} requires.
Namely, we achieve that all our suggestions actually make a difference for the SPARQL query.


\section{Efficient AC Queries}\label{sec:efficiency}

This section describes our main techniques to make AC queries efficient.
In our evaluation in Section \ref{sec:evaluation}, we impose a timeout for each AC query
(a user is only willing to wait so long for suggestions).
Efficiency is therefore a prerequisite for quality.


\subsection{Basic architecture of QLever, Virtuoso, and Blazegraph}\label{sec:efficiency:engines}

We implement the extensions described in the following subsections as extensions to {\QLever}.
In our evaluation in Section \ref{sec:evaluation}, we compare this to {\Virtuoso} and {\Blazegraph}.

{\QLever} \cite{DBLP:conf/cikm/BastB17} is an open-source query engine,
which stores the knowledge base triples in up to six permutations:
POS, PSO, SPO, SOP, OPS, OSP  (where S = subject, P = predicate, O = object).
The last four are optional and not needed for our evaluation.
%
Virtuoso \cite{OTHER:virtuoso} is a widely used SPARQL engine in research.
Virtuoso translates SPARQL queries to SQL queries, which are then executed by Virtuoso's own DB engine.
Triples are stored in one large table with three columns%
\footnote{There is another column for the ``graph name'', but in our evaluation,
we only use one ``graph'' (knowledge base) per Virtuoso instance}
(subject, predicate, object)
in column-layout with indexes PSO, POS, SP, and OP.
Blazegraph \cite{OTHER:blazegraph} is the query engine behind the official SPARQL endpoint
of the Wikidata Query Service \cite{OTHER:wdqs}.
Triples are stored in a B+ tree in three permutations: POS, SPO, and OSP.
There is little difference between the three engines in this respect.

In {\QLever}, each token (subject, predicate, object) is assigned a unique integer ID, stored as an 8-byte integer.
There are three types of tokens: IRIs, numerical literals, other literals.
For each type, the order of the IDs corresponds to the canonical
(lexicographical or numerical) order of the tokens of that type. 
The IDs form a contiguous interval starting from $0$ and the map from each ID to its token is stored in one large array.
Each permutation is stored on disk in a large array using $24 n$ bytes, where $n$ is the number of triples;
metadata is stored in memory.
SPARQL update operations are not supported, but building the index from scratch is very fast
(30 minutes for Fbeasy, 2.5 hours for Freebase, less than 24 hours for Wikidata,
on a single machine, as described in Section \ref{sec:evaluation:setup}).
At query time, each token from the query is translated to its ID using binary search on the above-mentioned map.
The query execution then takes place entirely in the ID space.
All intermediate results are fully materialized, as tables with fixed-size columns of size $8$ bytes each.
Only for the final result are the IDs converted to IRIs or literals again.
For the AC queries, there is always a {\LIMIT} $k$ for a small value of $k$, so that the last step takes negligible time.

In contrast, {\Virtuoso} and {\Blazegraph} are able to produce results ``one row at a time'',
without always having to fully materialize all intermediate results.
This can be an advantage for certain queries with a {\LIMIT},
but is a disadvantage for queries, where full materialization is required.
Most notably, this is typically required, when results are needed in a particular order.

\subsection{AC Queries for Predicates Using Patterns}\label{sec:patterns}

\def\calS{{\mathcal S}}
\def\calP{{\mathcal P}}
Our predicate AC queries use subqueries of the following kind.
\begin{sparqlq}
{\SELECT} ?entity ({\COUNT}({\DISTINCT} ?x) AS ?score) {\WHERE} \{ \\
\> {\context} ?x ?entity [] . \\
\} {\GROUPBY} ?entity {\ORDERBY} {\DESC}(?score)
\end{sparqlq}
As explained in Section \ref{sec:definition}, existing SPARQL engines materialize 
a table with one row for each triple of every \emph{?x}.
For a {\context} that constrains \emph{?x} little or not at all, this takes a (very) long time to compute.
To answer this kind of query efficiently, we make the following preprocessing:\\[1mm]
1. Let $\calS$ be the set of all distinct subjects in the knowledge base.
For each $x \in \calS$, compute the set of the distinct predicates from all triples that have $x$ as subject.
This set is called the (predicate) \emph{pattern} of $x$.
From these sets, compute the set $\calP$ of distinct patterns.
This is easily done by a pass over the above-mentioned SPO index.\\[1mm]
2. Give consecutive IDs to the patterns from $\calP$ and store the map from IDs to patterns in an array of size $|\calP|$.\\[1mm]
3. Store the map from each subject to its pattern ID in an array of size $|\calS|$.\\[1mm]
The following table provides statistics of this pre-processing for our three knowledge bases.
The fourth column counts the total size of the patterns,
where the size of a pattern is the number of predicates and each pattern is counted once.
The fifth column specifies the total memory consumption of the result of the pre-processing.

\def\Thr#1{\makebox[12mm][r]{$\mathbf{#1}$}}
\def\ThSubjects{\Thr{|\calS|}}
\def\ThPatterns{\Thr{|\calP|}}
\def\ThPredsinP{\Thr{\sum_{P\in\calP}|P|}}
\def\ThMemory{\makebox[12mm][r]{\bf Mem}}
\vspace{1mm}
\setlength{\tabcolsep}{2mm}
\begin{center}
{\renewcommand{\baselinestretch}{1.2}\normalsize
\noindent
\begin{tabular}{|l||r|r|r|r|}
\hline
                & \ThSubjects & \ThPatterns & \ThPredsinP & \ThMemory  \\\hline
{\bf Fbeasy}    &      $60\M$ &     $0.3\M$ &       $3\M$ &  $0.3\GB$  \\
{\bf Freebase}  &     $476\M$ &     $3.1\M$ &      $95\M$ &  $2.5\GB$  \\
{\bf Wikidata}	&    $2068\M$ &     $4.4\M$ &     $160\M$ &  $9.0\GB$  \\\hline
\end{tabular}}			
\end{center}
\vspace{1mm}

\noindent
To process the above query, we make use of these precomputed patterns as follows,
where steps 2 and 3 can be (and are) parallelized:\\[1mm]
1. Let $S \subseteq \calS$ be the set of subjects \emph{?x} from {\context} or $S = \calS$ if {\context} is empty.\\[1mm]
2. Look up the pattern IDs from all $x \in S$ in the precomputed array
and compute a map $c:{\mathcal P}_S \to \mathbb{N}$ that, for each pattern ID that occurs at least once,
counts how many $x \in S$ have that pattern ID.
This can be done in time linear in the size of $S$.\\[1mm]
3. For each pattern $P \in {\mathcal P}_S$, retrieve the corresponding set of predicate IDs and for
each $p$ in that set, increase a counter (initially $0$) by $c(P)$.
This takes time linear in $\sum_{P\in\calP_S}|P|$.\\[1mm]
4. Sort the encountered $p$ by the final counter values.
This yields the result for the query above.

\smallskip\noindent
The worst case for this algorithm is that
every subject has a different predicate pattern and exactly one triple for each predicate.
Then $|{\mathcal P}_S| = |S|$, each $c(P)$ is $1$, and Step 3 does exactly what the
naive algorithm described in Section \ref{sec:definition} would do.
However, in realistic knowledge bases, many subjects share the exact same set of predicates,
so that $|{\mathcal P}_s| < |S|$, and $\sum_{P\in\calP_S}|P|$ is much smaller than
the total number of triples of all $x \in S$.

For example, consider the AC query above for Fbeasy,
with {\context} $=$ \emph{?x $<$is\_a$>$ $<$Person$>$}.
In Fbeasy, there are $4.0\M$ persons with a total of $37\M$ triples.
They have only $|{\mathcal P}_s| = 115\K$ distinct patterns
with $\sum_{P\in\calP_S}|P| = 1.4\M$ predicates.
With {\QLever} extended by the pattern trick,
the query can be solved in under $0.1\s$.
With the standard query processing, {\QLever} takes $6.6\s$,
comprising $1.6\s$ for sorting $37\M$ elements.
%
%
%
%
%
%

We remark that {\Virtuoso} has a dedicated SP index, which maps each subject to its distinct predicates.
In theory, this index could be used to optimize the execution of the predicate AC queries.
However, despite our best efforts, we could not find a way to make
{\Virtuoso} use this index for queries like the above.

We briefly comment on how our patterns compare to the graph summary of Campinas {\etal}
\citep{DBLP:conf/dexaw/CampinasPCDT12} mentioned in Section \ref{sec:related:sparql}.
For the graph summary, entities that have the same set of types are grouped together.
For each group and each predicate, it is then precomputed how many entities of that group have that predicate.
For example, on Fbeasy, out of $3\,970\,825$ entities of type \emph{$<$Person$>$}, $103\,488$ have the predicate \emph{$<$award\_won$>$}.
This can be used for sensitive AC queries for predicates when the {\context} is exactly one triple that specifies the type.
Our preprocessing is more general and works for an arbitrary {\context}.

\subsection{Efficient names + prefix filtering}\label{sec:prefix_filtering}

Here is a variant of Example 1 from Section \ref{sec:definition} on Wikidata:
\begin{sparqlq}
{\SELECT} ?entity ?name (\COUNT(?x) {\AS} ?score) {\WHERE} \{\\
\> ?x wdt:P31 ?entity . \\
\> ?entity rdfs:label ?name . \\
\> {\FILTER} ({\LANG}(?name) = \verb|"|en\verb|"|) \\
\> {\FILTER} {\REGEX}(?name, "\verb|^|P")  \\
\} {\GROUPBY} ?entity ?name {\ORDERBY} DESC(?score)
\end{sparqlq}
A straightforward processing of this query has two problems.\\[1mm]
1. The \emph{rdfs:label} predicate has $359\M$ triples on Wikidata.
Materializing this table and then joining and filtering it is very slow.\\[.5mm]
2. The result of the first three rows of the {\WHERE} clause
is a table with $3$ columns and $72\M$ rows.
Materializing all \emph{?name} strings and applying a regex to each of them would be very slow.\\[1mm]
@1: Blazegraph's Wikidata setup circumvents this problem by providing a so-called \emph{label service}
via the SPARQL SERVICE keyword.
This label service provides labels in the desired languages via a separate process dedicated to only this task.
Our {\QLever} extension deals with this problem as follows:
duplicate all predicates with language literals during index building,
and split the duplicate into one predicate per language.
For the query above, then the predicate \emph{@en@rdfs:label} is used, which has $60\M$ rows.
The join with the first triple is then indeed materialized, but in ID space.\\[1mm]
@2: {\Virtuoso} and {\Blazegraph} do exactly this and are correspondingly slow.
Our {\QLever} extension makes use of the fact that literals also have IDs,
where the order of the IDs is the lexicographical order of the literals.
A {\FILTER} with a prefix regex like in the query above can then be realized with two binary searches.

\subsection{Caching and pinned results}\label{sec:caching}

We have extended {\QLever} by a thread-safe least-recently-used (LRU) query cache with the following features,
important for our AC queries.
The cache stores not only final results of a query, but also results from the intermediate operations
(which in SPARQL are always tables, too).
The query planner is aware of results in the cache:
the cost estimate for computing the result of a cached query is zero.
This is crucial for the processing of sequences of similar SPARQL queries,
as it naturally happens in our setting.

For example, assume we have typed the body of the
first query from the introduction (female oscar winners) until this point:
\begin{sparqlq}
\> ?subject \> $<$is\_a$>$   \>  <Person> \> . \\
\> ?subject \> $<$gender$>$  \>  <Female> \> . \\
\> ?subject \> {\cursor}
\end{sparqlq}
After the AC query for this predicate, the result for the first two triples (all female persons) are in the cache. 
Now assume that we have typed one triple further:
\begin{sparqlq}
\> ?subject \> $<$is\_a$>$      \>  <Person> \> . \\
\> ?subject \> $<$gender$>$     \>  <Female> \> . \\
\> ?subject \> $<$won award$>$  \>  ?award   \> . \\
\> ?award \> {\cursor}
\end{sparqlq}
To compute the result from the first three triples,
the best query plan is then to take the result from the first two triples from the cache
and join it with the result for the third triple.

Our cache also allows \emph{pinning} results.
These results will not be removed by an LRU eviction
(but there is a special command to clear the cache completely).
In our evaluation, we pin the results of the two building blocks {\entities} and {\predicates},
described in Section \ref{sec:ac_queries:building_blocks}, from which all our AC queries are built.
We store the results in two orders:
\emph{{\ORDERBY} ?entity} (for fast \emph{{\GROUPBY}} on \emph{?entity}) and \emph{{\ORDERBY} ?name} (for fast prefix search).
The memory consumption of these results is a small fraction of the index size
(for Wikidata: $6.7\GB$ for all pinned results).

\def\Unranked{Unranked}
\def\Agnostic{Agnostic}
\def\Sensitive{Sensitive}
\def\Mixed{Mixed}
\def\QLever{QLever}
\def\Blazegraph{Blazegraph}
\def\MRR#1{\mathrm{MRR}_#1}  
\def\MRRh#1#2#3#4{\makebox[5mm][r]{${#1}\%$}\makebox[17mm][r]{\tiny [1:#2\%, 2:#3\%, $\infty$:#4\%]}}  
\def\mb2#1{\makebox[10mm][r]{#1}}
\def\f#1{#1 {\scriptstyle\hspace{0.10em}\times\hspace{0.12em}} \infty}
\def\hlineg{\arrayrulecolor{lightgray}\hline}
\def\hlineb{\arrayrulecolor{black}\hline}
\def\WikidataPP{Wikidata$\rightleftharpoons$}

\section{Evaluation}\label{sec:evaluation}

In this section, we describe how we evaluated our approach,
and then present and discuss the results of this evaluation.
Materials for full reproducibility are available on our website: \url{http://ad.informatik.uni-freiburg.de/publications}.
In particular, the materials provide a web app that permit an interactive exploration of the
performance of the individual queries that are the basis of our main results (Table \ref{tab:main_results}).

\subsection{SPARQL Engines and Hardware}\label{sec:evaluation:setup}

We evaluate our own extension of {\QLever}, described in Section \ref{sec:efficiency},
against {\Virtuoso} and {\Blazegraph}, the basic architecture of which is
described in Section \ref{sec:efficiency:engines}.

All experiments were performed on a standard PC with an AMD Ryzen 7 3700X CPU (8 cores + SMT),
128 GB of DDR-4 RAM and 4 TB SSD storage (NVME, Raid 0).
We also ran our experiments on HDD storage (Raid 5), and found little difference.%
\footnote{However, indexing on HDD is much slower for {\Virtuoso} and {\Blazegraph},
but that is not the focus of this paper.}

{\QLever} was configured with a memory limit of $70\GB$ for query processing,
of which $30\GB$ were available to the query cache; see Section \ref{sec:caching}. 
Before each experiment, the query cache was cleared and the results of the queries
{\entities} and {\predicates} were pinned, as explained in Section \ref{sec:caching}.
For {\Virtuoso}, we use the latest release candidate of the open-source edition (7.2.6),
configured using the largest memory preset for $64\GB$ of RAM.%
\footnote{When scaling this preset up to $128\GB$ we found no significantly different results,
but frequently ran into problems with the out-of-memory killer.}
For {\Blazegraph}, we used the latest stable release (2.1.5),
configured according to {\Blazegraph}'s own recommendations for running Wikidata \cite{OTHER:blazegraph}.
In particular, {\Blazegraph} gets $16\GB$ for the JVM heap,
while the rest of the RAM is used for disk caching by the operating system.

We took great care to configure and use each engine optimally for the evaluation.
This comprises slight (equivalent) reformulations of the AC queries,
in order to avoid bad query plans; this is described in Section \ref{sec:evaluation:queries}.

\subsection{Knowledge Bases}\label{sec:evaluation:datasets}

We evaluate on the following three knowledge bases, already introduced in Section \ref{sec:introduction}.
We deliberately chose three knowledge bases with related content (general knowledge in this case),
but different sizes and combinations of human-readable vs.\ alpha-numeric IRIs.\\[1mm]
%
{\bf Fbeasy \cite{DBLP:conf/www/BastBBH14}}: $362\M$ triples, $50\M$ subjects, $2\K$ predicates.
All IRIs are simple and human-readable (e.g.\ \emph{$<$Meryl\_Streep$>$} or \emph{$<$gender$>$}).\\[.7mm]
{\bf Freebase \cite{DBLP:conf/sigmod/BollackerEPST08}}: $1.9\B$ triples, $125\M$ subjects, $785\K$ predicates.
Entity IRIs are alpha-numeric (e.g.\ \emph{fb:m.05dfkg3} for Meryl Streep), but most predicate IRIs are human-readable (e.g.\ \emph{fb:people.person.gender}).\\[.7mm]
{\bf Wikidata \cite{DBLP:journals/cacm/VrandecicK14}}: $6.9\B$ triples, $1.2\B$ subjects, $32\K$ predicates (complete dump from 01/2020).
Almost all IDs are alpha-numeric (e.g.\ \emph{wd:Q873} for Meryl Streep and \emph{wdt:P21} for gender).
To help {\Virtuoso} and {\Blazegraph}, we removed all non-English literals and all triples involving Wikimedia sitelinks.
The latter are only needed for the agnostic AC queries, which are executed solely by {\QLever};
see Section \ref{sec:evaluation:queries}.
In {\QLever}, we load the complete data ($11.3\B$ triples).
By the design of {\QLever}, this does not slow down queries (and doesn't make them faster either);
see Section \ref{sec:prefix_filtering}.


%
%
%
%

\begin{table*}[!ht]

\input{table-main-results.tex}

\vspace{3mm}
\caption{Query processing times and suggestion relevance for our three knowledge bases,
four completion modes, and three SPARQL engines.
For each to-be-completed token, three AC queries were issued, for prefix lengths 0, 3, and 7.
The column for the $\MRR{7}$ shows average results per AC query, and hence per prefix length.
The percentages in the ``Max'' column indicate the fraction of AC queries that timed out after $5\s$. 
For the $\MRR{7}$ and $\KS{7}$ those queries are treated as if the desired token appeared at position $\infty$
and the number of keystrokes required is the length of the token name plus $1$.
Column ``Sens'' shows the percentage of sensitive AC queries that did not time out.
See Sections \ref{sec:evaluation:datasets} - \ref{sec:evaluation:metrics} for more details,
and Section \ref{sec:evaluation:main_results} for a discussion.
For an interactive exploration of the raw data behind this table, see the web app
from the supplemental materials available for this paper on \url{http://ad.informatik.uni-freiburg.de/publications}.}
\label{tab:main_results}
\vspace{-4mm}
\end{table*}

\subsection{Autocompletion (AC) queries}\label{sec:evaluation:queries}

The basis for our evaluation are the 334 example queries from the Wikidata Query Service \cite{OTHER:wdqs}.
These queries cover a wide spectrum of SPARQL queries:
they range from simple to complex, use SPARQL features like UNION, OPTIONAL, MINUS, predicate paths, subqueries,
and cover the whole breadth of the content in the knowledge base.
We had to exclude some queries for technical reasons:\\[1mm]
{\bf Wikidata (301 queries):}
We excluded all 9 ``Lexeme'' queries because the respective triples are not part of the core Wikidata.
We excluded another 14 queries because they involve the SERVICE keyword as a crucial part of the query
and we did not want to make querying a remote SPARQL endpoint part of our evaluation.
We further excluded 10 queries involving distance computations or other mathematical expressions because those features are not implemented in {\QLever} and the queries became meaningless or impossible to compute without these computations. In several other queries we removed these expressions but were able to keep the queries.
For 3 queries, the first triple of the SPARQL query was of the form \emph{?x ?y ?z},
which is not a meaningful start in iterative query construction.
We moved it to the end of those queries.\\[1mm]
{\bf Freebase (115 queries):}
We manually translated those Wikidata queries, for which the contents are also contained in Freebase.
The translation is as close to the original Wikidata query as possible.
Note that Freebase became read-only in 2015 and Wikidata has much more contents by now
($6.9\B$ triples vs.\ $1.9\B$ triples, not counting literals in languages other than English).\\[1mm]
{\bf Fbeasy (99 queries):}
We manually translated all Freebase queries for which the contents are also contained in Fbeasy. 
Again, the translation is as close to the original query as possible.
Note that Fbeasy contains no $n$-ary information 
(for example: an award, the awardee, the work awarded, and the date). 
Instead, it retains only the binary ``core'' of each such tuple (for example: award and awardee).\\[1mm]
We consider all four modes of AC queries presented in Section \ref{sec:ac_queries}:
unranked, agnostic, sensitive, and mixed.
For a given mode and a given SPARQL query, we generate AC queries as follows:\\[1mm]
1. Consider each token (subject, predicate, or object) in the query that is either an IRI or a literal.
We exclude the special name predicates (\emph{fb:type.object.name} for Freebase,
\emph{rdfs:label} for Wikidata) because they occur in almost every query and are trivial to suggest
and would only distort our results.%
\footnote{In the $301$ Wikidata queries, there are $408$ triples of the form \emph{?x rdfs:label ?label}
(similarly for Freebase).
For entities with an \emph{rdfs:label}, that predicate is always among the most frequent suggestions.
There is only one \emph{rdfs:label} triple with a literal, namely \emph{?author rdfs:label "Ernest Hemingway"@en}.
The reason is that in SPARQL, one would directly use the entity (\emph{wd:Q23434})
rather than specifying it indirectly via its name.
With autocompletion, we easily find such an entity via (a prefix of) its name.}\\[1mm]
For each such token do the following:\\[1mm]
2. Compute {\context} as described in Section \ref{sec:ac_queries}
(only needed for the sensitive AC queries),
and depending on the position of the token, also determine {\subject} and {\predicate}.\\[1mm]
3. Choose a name from the result of the {\entities} query (canonical name and aliases)
for that token uniformly at random.\\[1mm]
4. From that name, compute three prefixes for {\prefix}: of length 0 (the empty word), 3, and 7.
For prefix lengths 3 and 7, if the name has less characters,
take {\prefix} as the complete name with $\$$ appended, indicating a full-word match.\\[1mm]
5. For each prefix length, pick the AC query template from Section \ref{sec:ac_queries}
according to the position of the token (subject, predicate, object)
and the mode (unranked, agnostic, sensitive, mixed).
Plug in {\context} and {\prefix}, and depending on the position also {\subject} and {\predicate}.\\[1mm]
For our first example query from the introduction (female Oscar winners),
this yields $7 \cdot 3 = 21$ AC queries per mode.
For our second example query (Oscars of Meryl Streep and corresponding films),
this yields $6 \cdot 3 = 18$ AC queries per mode.

The agnostic and the unranked AC queries only filter precomputed ranked lists
of entities and their names by the typed {\prefix}.
As explained in Section \ref{sec:ac_queries:agnostic}, this could be done with a special-purpose data structure.
In our evaluation, we evaluate them with {\QLever}, which turns out to be sufficiently fast;
see Table \ref{tab:main_results}.

We set the timeout for all AC queries (for all knowledge bases and all engines) to $5\s$.
In mixed mode, we launch a sensitive query and an agnostic query in parallel.
If the sensitive query does not finish within $1\s$, we take the result from the agnostic query.
That way, in mixed mode, we always obtain a result fast, it might just not be sensitive.

In our evaluation, we took great care to get the best query times for each engine, given its capabilities.
We found that for some of the AC queries, some engines produced suboptimal query plans,
leading to overly large processing times.
To remedy this, we used slightly different (but all equivalent) formulations of the AC queries for each engine.
The only exception:
in the predicate AC queries, we dropped the {\DISTINCT} for Virtuoso and Blazegraph
because almost all of their queries failed otherwise;
see the running times and explanations for Example 2 in Section \ref{sec:definition}.

The reader may wonder, why we did not evaluate AC queries after each keystroke.
We did this on purpose, for the following reason:
Ideally, a user does not have to type anything, and the desired token is suggested highly ranked already for prefix length $0$.
But if the suggestions for prefix length $0$ are not good, the user needs an idea of what to type anyway
and she might as well type a few letters instead of just one.
We chose $7$ as a representative for a prefix length that is not too long,
yet should sufficiently narrow down the search for most tokens.

\subsection{Evaluation metrics}\label{sec:evaluation:metrics}

We evaluate both objectives from our definition in Section \ref{sec:definition}.

\smallskip\noindent
\emph{Efficiency:} We report the percentage of AC queries that can be processed faster than $0.2\s$
(this feels close to instantaneous) and faster than $1.0\s$ (noticeable delay, but still acceptable).
If no query times out, we also report the maximum query time;
otherwise, we report the percentage of AC queries that timed out.

\smallskip\noindent
\emph{Relevance:} We report the percentage of sensitive queries that did not time out.
For these queries, \emph{all} suggestions are sensitive in the sense of our definition from Section \ref{sec:definition}.
Note that also an agnostic or unranked query can contain some sensitive suggestions.
However, this is of limited value to a user if they don't know which suggestions are sensitive and which are not.
What is important is the rank of the desired token, which we evaluate as follows.

We assume that suggestions are shown on ``pages'' of $k$ suggestions each.
Ideally, the desired token is on the first page (which is displayed after each keystroke).
In our evaluation, we take $k = 7$.
We use the following two metrics:

\smallskip\noindent
{\bf $\bm{\MRR{k}}$ (mean reciprocal rank):}
For each AC query, the reciprocal rank is $1/r$,
when $r$ is the index of the suggestion page on which the desired token occurs,
that is, at a position in $(r - 1)\cdot k \: .. \: r\cdot k - 1$, with the first position being $0$.
We report the mean reciprocal rank of all AC queries with a particular prefix length (0, 3, and 7).
The maximum value of $\MRR{7}$ is $100\%$; it is achieved when each token appears on the first page.

Note that the reciprocal rank is a very natural measure in our setting:
we only have one relevant item and the ``gain'' for the user
indeed decreases sharply with the index of the page where the item occurs.
A user would rather continue typing instead of scrolling down much further in the list of suggestions.

\smallskip\noindent
{\bf $\bm{\KS{k}}$ (number of keystrokes):}
For each token, the number of keystrokes is the minimal prefix length (out of $0$, $3$, and $7$),
for which the token appears on the first page of suggestions.
If it is not on the first page even for prefix length $7$,
we take the number of keystrokes for that token as the length of the name of the token plus 1. 
This corresponds to typing the full name and indicating that it is not a prefix, but the full name.

\subsection{Main results and discussion}\label{sec:evaluation:main_results}

Table \ref{tab:main_results} summarizes our main results.
It contains a lot of information, which we tried to arrange as clearly as possible.
The main takeaways are as follows.
The table shows only results for those AC queries
for which the result is not already fully precomputed;
these are discussed separately at the end of this section.

\def\iskip{\\[1mm]}
\def\header#1{\vspace{1mm}\noindent{\bf #1}}

\header{Sensitive AC queries help relevance a lot.}
Compare the $\MRR{7}$ of {\Agnostic} and {\Sensitive} using {\QLever} on Wikidata.
The values at prefix length 0 are $6\p$ vs. $50\p$.
This shows that without typing anything,
the desired token is hardly ever on the first pages of suggestions with {\Agnostic},
but frequently on the first or second page for {\Sensitive}.
This case is particularly important because if you have to type something,
then you already need an idea what you are looking for.
In Section \ref{sec:definition}, Example 3 showed a typical situation,
where it is very hard to guess even the first few letters of the desired token.
After typing three letters, the result is almost always on the first page for {\Sensitive},
and {\Agnostic} is also becoming better.

We also see a clear difference in the $\KS{7}$,
that is, with {\Sensitive} a user has to type considerably less to get the desired token
on the first page of suggestions.
However, we consider the $\MRR{7}$ broken down by prefix length to be the more important and insightful measure.
%
The results on the smaller knowledge bases are similar, though the difference is less dramatic.

\header{Sensitive AC queries are feasible with {\QLever}, but often fail with {\Virtuoso} and {\Blazegraph}.}
Even on the very large Wikidata, {\QLever} can provide $90\p$ of the sensitive suggestions in under a second,
and $71\p$ under $200\ms$.
Only few AC queries time out after $5\s$ on Wikidata, and hardly any on the smaller knowledge bases.
On Wikidata, two thirds of the timed out AC queries are object queries
involving predicate paths with huge intermediate results.
The most frequently occurring is \emph{wdt:P31/wdt:P279*}
(``instance of'' followed by one or more ``subclass of''),
which matches $1.5\B$ triples.
In Section \ref{sec:conclusions}, we discuss an ad-hoc fix for this problem.
For our evaluation, we explicitly wanted an approach with minimal configuration
and no ad-hoc adjustments.

For {\Virtuoso}, a quarter of the AC queries time out.
For {\Blazegraph}, over half of the AC queries time out and hardly any are fast.
For the smaller knowledge bases, {\QLever} has practically no timeouts,
but {\Virtuoso} and {\Blazegraph} still have many.
Note that $5\s$ is a rather generous timeout:
users hardly want to wait that long for a suggestion in an interactive scenario.
In our own experience, usage becomes annoying if you have to wait beyond $1\s$.

\header{{\Agnostic} AC queries are always fast;
relevance is bad for prefix length $\mathbf{0}$ but quite good for longer prefix lengths.}
All agnostic AC queries can be processed in well under one second.
This is not surprising, since the results are essentially precomputed.
The only non-trivial work to do at query time
is to filter the precomputed results by the prefix typed.
As explained in Section \ref{sec:ac_queries:agnostic},
this could also be achieved with a special-purpose data structure.
But it is nice that we don't need such a data structure,
but can just use our SPARQL engine for it.

For an agnostic AC query with prefix length $0$,
the desired token will be rarely among the top suggestions
because of the complete lack of contextual information.
But a prefix length of $3$ or even $7$ is often enough to restrict the suggestions sufficiently,
even without {\context}.
This is important in order to understand the results for the mixed AC queries, discussed below.

{\Virtuoso} and {\Blazegraph} both perform very poorly for agnostic AC queries,
which is why we do not report them in Table \ref{tab:main_results}.
The reason is that both engines handle prefix searches on large lists of strings
very inefficiently.
See the discussion and running time in Example 1 in Section \ref{sec:definition}
and in Section \ref{sec:prefix_filtering}.

\header{{\Mixed} AC queries are a good compromise between sensitivity and performance}
{\Mixed} always produces a result within $1\s$ and so never times out.
The reasons is that agnostic AC queries can always be processed in under $1\s$. 
{\Mixed} is therefore a perfect solution with respect to efficiency.

The $\MRR{7}$ of {\Mixed} for the important prefix-length-$0$ case 
comes very close to the $\MRR{7}$ for {\Sensitive}.
For prefix lengths 3 and 7, {\Mixed} is even better,
the difference being larger for {\Virtuoso} and {\Blazegraph}.
The reason is that when the sensitive AC query times out,
the reciprocal rank (RR) for {\Sensitive} is $0$ because we don't have any suggestions.
With {\Agnostic}, we get at least some suggestions.
As discussed for {\Agnostic} above, these suggestions are not very good for prefix length $0$,
but quite good for prefix lengths 3 and 7.
For prefix length $0$, the $\MRR{7}$ of {\Mixed} is worse than that of {\Sensitive}
because of the smaller timeout ($1\s$ instead of $5\s$).

A downside of {\Mixed} is that for the agnostic suggestions, even when they are good,
the user does not know if they are good.
A user interface could indicate this fact by showing suggestions from agnostic queries 
in a different color or with some other visual marker.
On Wikidata using {\QLever}, $88\%$ of the mixed AC queries are sensitive.
Note that the number is smaller than in the {\Sensitive} row because of the smaller timeout.

\header{{\Unranked} AC queries perform very poorly on large knowledge bases.}
Recall that the suggestions of {\Unranked} are the same as those of {\Agnostic},
but without ranking them by score.
We include this mode in our evaluation to show how important ranking is.
On Wikidata, even for a prefix length of $3$, the relevance of {\Unranked} is very poor ($\MRR{7} = 8\%$).
For a prefix length of $7$, the $\MRR{7}$ rises to $53\p$, but it's still much worse than the $92\%$ of {\Agnostic}.
Note that ranking for sensitive autocompletion is mainly an efficiency problem:
often very large amounts of suggestions have to be computed and sorted.
In some of the previous work we discussed, ranking was omitted due to this reason; see Section \ref{sec:related}.

\header{Results for AC queries without context.}
Table \ref{tab:main_results} does not include predicate AC queries
with a variable {\subject} and empty {\context}
or for subject AC queries.
The reason is that the suggestions for these queries are exactly the results of
our building blocks {\predicates} and {\entities}, which we precompute once and then pin to the cache.
These queries can hence be processed just as fast as the agnostic AC queries.

There are $112$ such tokens for Fbeasy, $132$ for Freebase, and $363$ for Wikidata;
so about one fourth of all tokens for Fbeasy and Freebase, and one fifth for Wikidata
(each token corresponds to three AC queries, one for each prefix length).
The average $\MRR{7}$ per prefix length (0-3-7) for these AC queries are:
$59\p$-$95\p$-$99\p$ for Fbeasy, $52\p$-$97\p$-$99\p$ for Freebase, and $21\p$-$78\p$-$95\p$ for Wikidata.
The $\KS{7}$ results are: $1.94$ for Fbeasy, $1.90$ for Freebase, $5.00$ for Wikidata.

These $\MRR{7}$ and $\KS{7}$ results (without context) are significantly better than
for the agnostic queries in Table \ref{tab:main_results} (with context).
There are two reasons for this.
First, most of the AC queries without context are predicate suggestions,
and there are much fewer predicates than entities, so that it is easier to have a desired predicate highly ranked.
Second, the predicates in a SPARQL query that have no context are often 
frequent predicates, like \emph{$<$is\_a$>$} on Fbeasy or \emph{wdt:P31} (instance of) on Wikidata.

\section{Conclusions}\label{sec:conclusions}

We showed how to perform context-sensitive SPARQL autocompletion
with very good relevance and efficiency,
for a large variety of queries on three different knowledge bases.
All suggestions were themselves provided via SPARQL queries,
on the same knowledge base on which we want to construct SPARQL queries
with the aid of autocompletion.
That way, our scheme can be used with any (standard-conforming) SPARQL engine.

We showed that on very large knowledge bases (like Wikidata),
many autocompletion queries are hard for existing SPARQL engines.
We showed two ways out.
First, we showed how to extend an existing open-source SPARQL engine
to deal with most of these hard queries efficiently.
Second, we introduced a mixed mode that sacrifices context-sensitivity for efficiency.
Here are some interesting directions for future work:\\[1mm]
1. An obvious next step is to implement a user interface based on the autocompletion mechanism described in this paper.
Such a user interface would also have to implement various syntactic suggestions; see Section \ref{sec:introduction:scope}. While this is straightforward algorithmically, it poses various challenge from a UX perspective.\\[1mm]
2. Our pattern preprocessing, described in Section \ref{sec:patterns},
goes a long way towards providing predicate suggestions in interactive time.
There is still room for improvement, however.
Especially on Wikidata, there are large groups of entities, which have very similar but not exactly equal patterns.
If we could identify these groups, we would have even fewer patterns and could store for each entity
the common pattern and its small difference to that pattern.
This has the potential to speed up predicate suggestions further.\\[1mm]
3. Our autocompletion is a fantastic help for finding the individual IRIs in a SPARQL query.
However, for some SPARQL queries, it can be hard to know in advance which information
is represented as a predicate and which information is represented as an object in the knowledge base.
For example, when looking for US presidents in Wikidata,
the right predicate is \emph{wd:P39} (position held) and the right object is \emph{wd:Q11696} (President of the US).
But the predicate might as well have been \emph{wdt:P31} (instance of)
or the information about the job title could have been split over two triples (being president and of which country).
A more intelligent autocompletion might provide suggestions for both predicates and objects
and figure out automatically where to place them in the query.\\[1mm]
%
4. Among the remaining very hard AC queries in our evaluation are 
queries involving predicate paths like \emph{wdt:P31/wdt:P279*} (instance of, subclass of).
A simple ad-hoc fix for this problem is as follows:
Identify all predicates in the knowledge base, which can meaningfully form such chains.
Wikidata even provides a dedicated predicate (\emph{wdt:P6609}, named ``value hierarchy property'') for this
(for example, it connects \emph{wdt:P31} with \emph{wdt:P279}, and \emph{wdt:P279} with itself).
For each of these paths, precompute the objects ordered by frequency.
These queries are hard to compute, but the result size is on the same order as for our {\entities} query.
For the corresponding AC query, then show the precomputed result.
This is not fully sensitive, but close to it:
for example, in the case of \emph{wdt:P31/wdt:P279*}, only ``types'' are suggested,
instead of entities that obviously make no sense as objects of this predicate path.\\[1mm]
%

\vspace*{-1mm}
\bibliographystyle{ACM-Reference-Format}
\bibliography{sparql-autocompletion}

\end{document}

%% file: related-work.tex
\def\etal{et al.}
\section{Related Work}\label{sec:related}

We first discuss related work on SPARQL autocompletion via queries in (variants of) SPARQL.
We then provide a brief overview of other approaches assisting users to formulate their query.

\subsection{SPARQL autocompletion via SPARQL}\label{sec:related:sparql}

Campinas {\etal} \citep{DBLP:conf/dexaw/CampinasPCDT12} present an autocompletion system that is able to recommend predicates and types (i.e. objects of a \emph{type} predicate).
This approach uses AC queries, but runs them on a smaller \emph{graph summary} that only captures the structure of the data,
which helps efficiency, but harms relevance.
The idea behind the graph summary is similar to our pattern trick in Section \ref{sec:patterns}
but with the important difference, that our implementation does not affect relevance;
we come back to this in Section \ref{sec:patterns}.
In a follow-up paper, Campinas \citep{DBLP:conf/semweb/Campinas14} presents a system called \emph{Gosparqled} 
that uses AC queries similar to ours.
However, suggestions are again only possible for predicates and types and the results are not ranked.
We briefly discuss the two objectives from our definition in Section \ref{sec:definition} for these papers.\\[.5mm]
\noindent\emph{Relevance:}
The AC queries of Gosparqled are similar to those from Section \ref{sec:definition},
but with a {\LIMIT} (of 10, 100 or 500).
Hence all suggestions are meaningful continuations of the part of the query already typed.
In the evaluation, the suggestions are compared to suggestions without this {\LIMIT}.
The average Jaccard similarity varies between $11\%$ and $62\%$,
depending on the complexity of the query and the {\LIMIT}.
With the graph summary, the Jaccard similarity does not decrease significantly.
Our sensitive AC queries impose no {\LIMIT} and, by definition, achieve a similarity of $100\%.$\\[.5mm]
%
\noindent\emph{Efficiency:} 
Gosparqled achieves response time below $0.2\s$ for $94\%$ of the AC queries.
The main reason is the mentioned {\LIMIT}, which severely impacts relevance.
The graph summary reduces the average run time further by about $35\%$. 
The knowledge base used for the evaluation consists of less than half the amount of triples than our smallest dataset, Fbeasy.
The evaluation setup is also different in that for each suggestion,
only the desired token is removed from the query.
That way, the AC queries have very restricting contexts, which helps efficiency a lot.
In contrast, we simulate typing the query from beginning to end,
which results in many very hard AC queries that have to deal with huge intermediate results.

Jarrar and Dikaiakos \citep{DBLP:journals/tkde/JarrarD12} present autocompletion for \emph{MashQL}, a variant of SPARQL with essentially the same functionality as SPARQL.
They also use AC queries to suggest entities to the user.
MashQL's completions are only context-sensitive for \emph{linear-shaped} queries.
For example, if the user has already typed \emph{?x1 <place of birth> ?x2 . ?x2 <country> ?x3 . ?x3 \cursor },
MashQL will consider this context.
But if the user has typed \emph{?x <place of birth> <Berlin> . ?x <gender> <Female> . ?x \cursor },
MashQL will make the same suggestion for each predicate, without taking the previous context into account.
To be able to run the AC queries more efficiently, they use two graph summaries.
In one graph summary, entities with the same \emph{outgoing} paths are grouped together and in the other one,
entities with the same \emph{incoming} paths are grouped together.
These summaries only lead to correct results for the above mentioned \emph{linear-shaped} queries.

Bast {\etal} \citep{bast2012broccoli} present a system called \emph{Broccoli},
which provides context-sensitive suggestions for tree-shaped queries
and depicts the queries as trees.
The underlying query language is equivalent to SPARQL, restricted to trees and basic graph patterns.
The focus of the paper is on extending the query language by a text-search component
and on providing efficient autocompletion for this component as well.
The user interface and the evaluation work with a knowledge base with human-readable IRIs
(similar to Fbeasy, but only 26M triples) and do not support synonyms or aliases.

Ferré \citep{DBLP:journals/semweb/Ferre17} 
presents a system called \emph{SPARKLIS}, which suggests context-sensitive continuations for SPARQL queries.
Queries are formulated in natural language (for instance
``Give me every person whose gender is female and who won an award that is an Oscar'').
It supports most of the functionality of SPARQL. 
Suggestions are obtained via AC queries similar to those from Section \ref{sec:definition}. 
In order to address efficiency issues, there is a {\LIMIT} on the results and results are not ranked. 
As the user types, results of the partially written query are computed and shown.
These intermediate results are used to compute the suggestions faster
(similar to our cache described in Section \ref{sec:caching}).

\subsection{Other assisted formulation systems}\label{sec:related:other}

There is a wide literature on other approaches to assist the user in creating SPARQL queries
or get results from a knowledge base.
We provide an overview by briefly describing a representative system for each approach.


Arenas {\etal} \citep{DBLP:journals/ws/ArenasGKMZ16} present \emph{SemFacet},
a faceted search system for RDF data.
The user starts with a keyword query (e.g. ``person''),
from which the system computes a list of matching entities,
For each predicate, users can narrow down the results by choosing specific objects (e.g. \emph{<Actor>}).
The suggested predicates and objects are not ranked, for efficiency reasons.
The underlying queries are SPARQL queries, restricted to trees.
%
One of the first systems of this kind is \emph{BrowseRDF} by Oren {\etal} \citep{DBLP:conf/semweb/OrenDD06}.
BrowseRDF shows entities for the root variable, while in SemFacet any variable can be selected.
BrowseRDF also suggests inverse predicates and negation.



Rafes {\etal} \citep{DBLP:conf/eScience/RafesABR18} introduce a log-based system called \emph{SPARQLets-Finder},
which takes a basic graph pattern (essentially: a part of a SPARQL query)
and suggests a ranked lists of graph patterns to extend the query.
The suggestions are based on a hierarchical clustering of the graph patterns in the given query log,
using a metric introduced in the paper.
The evaluation compares the recommendations with those made by a related tool.


Khoussainova {\etal} \citep{DBLP:journals/pvldb/KhoussainovaKBS11} present \emph{SnipSuggest},
a log-based autocompletion tool for SQL.
The user starts with a simple query.
The system then suggests additional \emph{features} to the user,
for example, a table in the {\FROM} clause or a condition in the {\WHERE} clause.
SnipSuggest achieves this by transforming the partially written query into a set of such features,
and then computing the most popular additional features according to the query log.
Alternatively, the suggestions are optimized for diversity.
The evaluation shows that the suggestions are useful and
better than from naive approaches, which recommend the overall most popular features.


Bast {\etal} \citep{bast2015more} present \emph{Aqqu}, a state-of-the-art system for \emph{semantic parsing}.
The task is to translate a given question in natural language into an equivalent SPARQL query.
The translation is learned only from question-answer pairs (no ground truth SPARQL queries are needed).
There is a wide literature on this setting.


Arkoudas and Yahya \citep{DBLP:conf/cikm/ArkoudasY19} also present
a system that translates questions in natural language to structured queries (SPARQL or SQL).
The system is interactive:
as the user types the question, the system suggests possible continuations.
Three different suggestion algorithms are combined:
Two of these algorithms are \emph{log-based}
(one returns suggestions when everything typed so far matches, the other also supports partial matches).
The third is based on templates. 
That way, the system makes effective use of a query log but also covers queries unseen so far.


Lehmann and Bühmann \citep{DBLP:conf/esws/LehmannB11} present \emph{AutoSPARQL},
which suggests whole SPARQL queries (restricted to trees) in a back-and-forth with the user.
The user starts by formulating a simple query that contains at least one result of the target query.
AutoSPARQL then asks whether certain results should be contained in the target query or not.
Based on the user feedback, new SPARQL queries are suggested and new feedback is asked.
This is repeated until the desired query is found or no tree query exists.
In the evaluation, on average 5 iterations were needed to find the target query or conclude that no matching query exists.


Fan {\etal} \citep{DBLP:conf/icde/FanLZ11} present \emph{SQLSUGG},
which uses keyword search to suggest entire queries to the user.
Keywords may refer to either table values, the meta-data (e.g. names of relation tables or attributes), or aggregate functions (e.g. the function {\COUNT}).
To get suggestions, they use templates and rank them  based on the keywords typed.

%% file: table-main-results.tex

\setlength{\tabcolsep}{2.5mm}
\def\spa{\hspace{0.7mm}}
\def\spb{\hspace{2.0mm}}
\def\b#1{{\color{darkblue} #1}}
\def\bb#1{{\color{darkblue}\boldmath\textbf{#1}}}
{\renewcommand{\baselinestretch}{1.2}\normalsize
\begin{tabular}{|l|l||r|r|r@{\spa}r@{\spb}r@{\spa}r@{\spb}r@{\spa}r||r@{\spa}|r@{\spa}r@{\spb}r@{\spa}r@{\spb}r@{\spa}r|r|}

\hline
\multicolumn{2}{|l||}{\textbf{Fbeasy (314 tokens)}} & $\mathbf{\le 0.2\s}$ & $\mathbf{\le 1.0\s}$ & \multicolumn{6}{r||}{\boldmath\textbf{Max}} & $\mathbf{Sens}$ & \multicolumn{6}{r|}{\boldmath\textbf{MRR$_7$}} & \boldmath\textbf{KS$_7$} \\\hline
Unranked & Qlever & $100\p$ & $100\p$ & \multicolumn{6}{r||}{$444\ms$} & $0\p$ & 0: & $0\p$ & 3: & $33\p$ & 7: & $72\p$ & 8.04 \\
Agnostic & Qlever & $100\p$ & $100\p$ & \multicolumn{6}{r||}{$470\ms$} & $0\p$ & 0: & $25\p$ & 3: & $86\p$ & 7: & $96\p$ & 3.75 \\
\hline
Sensitive & Blazegraph & $26\p$ & $42\p$ & \multicolumn{6}{r||}{$45\p > 5\s$} & $55\p$ & 0: & $38\p$ & 3: & $53\p$ & 7: & $53\p$ & 5.74 \\
Sensitive & Virtuoso & $37\p$ & $61\p$ & \multicolumn{6}{r||}{$25\p > 5\s$} & $75\p$ & 0: & $58\p$ & 3: & $65\p$ & 7: & $65\p$ & 3.79 \\
\b{Sensitive} & \b{Qlever} & \b{$91\p$} & \b{$97\p$} & \multicolumn{6}{r||}{\b{$1\p > 5\s$}} & \b{$99\p$} & \b{0:} & \b{$67\p$} & \b{3:} & \b{$96\p$} & \b{7:} & \b{$97\p$} & \b{1.77} \\
\hline
Mixed & Blazegraph & $26\p$ & $97\p$ & \multicolumn{6}{r||}{$1234\ms$} & $45\p$ & 0: & $50\p$ & 3: & $94\p$ & 7: & $97\p$ & 2.52 \\
Mixed & Virtuoso & $47\p$ & $100\p$ & \multicolumn{6}{r||}{$1000\ms$} & $72\p$ & 0: & $57\p$ & 3: & $96\p$ & 7: & $91\p$ & 2.07 \\
\bb{Mixed} & \bb{Qlever} & \bb{$90\p$} & \bb{$100\p$} & \multicolumn{6}{r||}{\bb{$1000\ms$}} & \bb{$98\p$} & \bb{0:} & \bb{$66\p$} & \bb{3:} & \bb{$96\p$} & \bb{7:} & \bb{$98\p$} & \bb{1.75} \\
\hline

\multicolumn{17}{l}{} \\[-2mm]
\hline
\multicolumn{2}{|l||}{\textbf{Freebase (478 tokens)}} & $\mathbf{\le 0.2\s}$ & $\mathbf{\le 1.0\s}$ & \multicolumn{6}{r||}{\boldmath\textbf{Max}} & $\mathbf{Sens}$ & \multicolumn{6}{r|}{\boldmath\textbf{MRR$_7$}} & \boldmath\textbf{KS$_7$} \\\hline
Unranked & Qlever & $100\p$ & $100\p$ & \multicolumn{6}{r||}{$645\ms$} & $0\p$ & 0: & $0\p$ & 3: & $24\p$ & 7: & $55\p$ & 9.05 \\
Agnostic & Qlever & $100\p$ & $100\p$ & \multicolumn{6}{r||}{$621\ms$} & $0\p$ & 0: & $12\p$ & 3: & $83\p$ & 7: & $94\p$ & 4.52 \\
\hline
Sensitive & Blazegraph & $29\p$ & $46\p$ & \multicolumn{6}{r||}{$41\p > 5\s$} & $59\p$ & 0: & $39\p$ & 3: & $57\p$ & 7: & $58\p$ & 5.32 \\
Sensitive & Virtuoso & $41\p$ & $59\p$ & \multicolumn{6}{r||}{$19\p > 5\s$} & $81\p$ & 0: & $43\p$ & 3: & $79\p$ & 7: & $80\p$ & 3.79 \\
\b{Sensitive} & \b{Qlever} & \b{$87\p$} & \b{$97\p$} & \multicolumn{6}{r||}{\b{$0.5\p > 5\s$}} & \b{$100\p$} & \b{0:} & \b{$60\p$} & \b{3:} & \b{$97\p$} & \b{7:} & \b{$99\p$} & \b{2.06} \\
\hline
Mixed & Blazegraph & $28\p$ & $100\p$ & \multicolumn{6}{r||}{$1041\ms$} & $49\p$ & 0: & $42\p$ & 3: & $93\p$ & 7: & $98\p$ & 2.68 \\
Mixed & Virtuoso & $42\p$ & $100\p$ & \multicolumn{6}{r||}{$1000\ms$} & $62\p$ & 0: & $43\p$ & 3: & $97\p$ & 7: & $99\p$ & 2.42 \\
\bb{Mixed} & \bb{Qlever} & \bb{$73\p$} & \bb{$100\p$} & \multicolumn{6}{r||}{\bb{$1000\ms$}} & \bb{$82\p$} & \bb{0:} & \bb{$59\p$} & \bb{3:} & \bb{$96\p$} & \bb{7:} & \bb{$99\p$} & \bb{2.15} \\
\hline

\multicolumn{17}{l}{} \\[-2mm]
\hline
\multicolumn{2}{|l||}{\textbf{Wikidata (1258 tokens)}} & $\mathbf{\le 0.2\s}$ & $\mathbf{\le 1.0\s}$ & \multicolumn{6}{r||}{\boldmath\textbf{Max}} & $\mathbf{Sens}$ & \multicolumn{6}{r|}{\boldmath\textbf{MRR$_7$}} & \boldmath\textbf{KS$_7$} \\\hline
Unranked & Qlever & $100\p$ & $100\p$ & \multicolumn{6}{r||}{$635\ms$} & $0\p$ & 0: & $0\p$ & 3: & $8\p$ & 7: & $53\p$ & 10.87 \\
Agnostic & Qlever & $100\p$ & $100\p$ & \multicolumn{6}{r||}{$696\ms$} & $0\p$ & 0: & $6\p$ & 3: & $64\p$ & 7: & $92\p$ & 5.88 \\
\hline
Sensitive & Blazegraph & $3\p$ & $27\p$ & \multicolumn{6}{r||}{$59\p > 5\s$} & $41\p$ & 0: & $26\p$ & 3: & $36\p$ & 7: & $37\p$ & 7.93 \\
Sensitive & Virtuoso & $35\p$ & $53\p$ & \multicolumn{6}{r||}{$24\p > 5\s$} & $76\p$ & 0: & $38\p$ & 3: & $67\p$ & 7: & $67\p$ & 5.26 \\
\b{Sensitive} & \b{Qlever} & \b{$71\p$} & \b{$90\p$} & \multicolumn{6}{r||}{\b{$6\p > 5\s$}} & \b{$94\p$} & \b{0:} & \b{$50\p$} & \b{3:} & \b{$92\p$} & \b{7:} & \b{$95\p$} & \b{2.91} \\
\hline
Mixed & Blazegraph & $0\p$ & $98\p$ & \multicolumn{6}{r||}{$1066\ms$} & $25\p$ & 0: & $22\p$ & 3: & $71\p$ & 7: & $94\p$ & 4.71 \\
Mixed & Virtuoso & $35\p$ & $100\p$ & \multicolumn{6}{r||}{$1002\ms$} & $59\p$ & 0: & $36\p$ & 3: & $76\p$ & 7: & $91\p$ & 4.22 \\
\bb{Mixed} & \bb{Qlever} & \bb{$68\p$} & \bb{$100\p$} & \multicolumn{6}{r||}{\bb{$1000\ms$}} & \bb{$88\p$} & \bb{0:} & \bb{$47\p$} & \bb{3:} & \bb{$93\p$} & \bb{7:} & \bb{$98\p$} & \bb{2.76} \\
\hline

\end{tabular}}